\documentclass[a4paper,10pt]{article}

\usepackage[cp1252]{inputenc}
\usepackage[english]{babel}
\usepackage{amsmath,amsfonts,amssymb,amsthm,amstext}
\usepackage{longtable,cite,url,graphicx}
\usepackage{color,relsize}

\newcommand\mysectiona[2]{%
\section
[#1]
{#2}
}

\newcommand\mysectionb[2]{%
\subsection
[#1]
{#2}
}

\newcommand{\notioninform}[1]{\text{\bf #1}}

\newcommand{\stdsetnameinform}[1]{%
\mathbb{#1}}

\newcommand{\classname}[1]{%
{\bf #1}}

\newcommand{\classnameinform}[1]{%
\text{\bf #1}}

\newtheorem{definition}{Definition}
\newtheorem{proposition}{Proposition}

\newtheorem{theorem}{Theorem}

\newtheorem*{notation*}{Notation}
\newtheorem*{definition*}{Definition}
\newtheorem*{proposition*}{Proposition}
\newtheorem*{lemma*}{Lemma}
\newtheorem*{theorem*}{Theorem}
\newtheorem*{corollary*}{Corollary}

\begin{document}

%/////////////////////////////////////////////////////////////////////////////
%
\title{
Computable real function $\mathcal{F}$\\
such that $\mathcal{F}\in\notioninform{FEXPTIME}_{C[0,1]}$\\
but $\mathcal{F}\notin\notioninform{FPTIME}_{C[0,1]}$
}
\author{
Sergey\,V.\,Yakhontov
\date{}}

{\def\thefootnote{}
\footnotetext{
\centering
Sergey\,V.\,Yakhontov: Ph.D. in Theoretical Computer Science, Dept. of Computer Science, Faculty of\\
Mathematics and Mechanics, Saint Petersburg State University, Saint Petersburg, Russian Federation;\\
e-mail: SergeyV.Yakhontov@gmail.com, S.Yakhontov@spbu.ru; phone: +7-911-966-84-30;\\
personal Web page: \url{https://sites.google.com/site/sergeyvyakhontov/}; 25 Apr 2014}
%
%/////////////////////////////////////////////////////////////////////////////
\maketitle

%///////////////////////////////////////////////////////

\abstract{
%/////////////////////////////////////////////////////////////////////////////
%
In the present paper, a computable real function $\mathcal{F}$ on $[0,1]$ is constructed
such that there exists an exponential time algorithm for the evaluation of the function
on $[0,1]$ on Turing machine but there does not exist any polynomial time
algorithm for the evaluation of the function on $[0,1]$ on Turing machine (moreover,
it holds for any rational point on $(0,1)$).
%
%/////////////////////////////////////////////////////////////////////////////
}

\vspace{0.2cm}
%/////////////////////////////////////////////////////////////////////////////
%

\tableofcontents

%/////////////////////////////////////////////////////////////////////////////
%
\mysectiona
{Introduction}
{Introduction}
In the present paper, we consider computable real numbers and functions that are
represented by Cauchy functions computable on Turing machines \cite{K91}
(main results regarding computational complexity of computations on Turing machines
can be found in \cite{DK00}).

Book \cite{K91} studies mainly polynomial time computable real numbers and functions.
It is a common point of view that polynomial time algorithms can be considered as efficient
algorithms, so such algorithms are very important for theoretical and practical computer science.

In book \cite{K91} a general theory of polynomial time computable real numbers and functions
is considered. But it is interesting to consider specific examples of computable real numbers
and functions with specific lower and upper bounds of time and space complexity of them. 

In the present paper, a computable real function $\mathcal{F}$ on $[0,1]$ is constructed
such that the function is exponential time computable on $[0,1]$ but it is not polynomial time
computable on $[0,1]$ (moreover, it holds for any rational point on $(0,1)$).

To show that function $\mathcal{F}$ is not polynomial time computable on $[0,1]$, it is proved
that there does not exist any polynomial time modulus function of uniform continuity of function
$\mathcal{F}$ on $[0,1]$ (moreover, there does not exist any polynomial time modulus function of
continuity of function $\mathcal{F}$ at any rational point on $(0,1)$). So, a function such
that there is no efficient algorithm to evaluate it is given.

The construction of function $\mathcal{F}$ is very similar to the constructions
of nowhere-differentiable real functions like Weierstrass function
(many of such functions could be found in \cite{T03}) in the sense that
function $\mathcal{F}$ is an infinite sum of simple functions with certain
properties. A similar construction can also be found in \cite{K91} (in the examples
regarding complexity of derivatives of polynomial time computable real functions).
%
%/////////////////////////////////////////////////////////////////////////////
%/////////////////////////////////////////////////////////////////////////////
%
\mysectionb
{$CF$ computable real numbers and functions}
{$CF$ computable real numbers and functions}
Cauchy functions in the model defined in \cite{K91} are functions binary converging to real numbers.
A function $\phi:\mathbb{N}\rightarrow\mathbf{D}$ (here $D$ is the set of dyadic rational
numbers) is said to binary converge to real number $x$ if $$|\phi(n)-x|\leq 2^{-n}$$ for all $n\in\mathbb{N}$;
$CF_x$ denotes the set of all functions binary converging to $x$.

Real number $x$ is said to be a $CF$ computable real number if
$CF_x$ contains a computable function $\phi$.

Real function $f$ on $[a,b]$ is said to be a $CF$ computable function on $[a,b]$
if there exists a function-oracle Turing machine $M$ such that for all $x\in[a,b]$
and for all $\phi\in CF_{x}$ function $\psi$ computed by $M$ with oracle $\phi$
is in $CF_{f(x)}$.

When the time complexity of a computable function $f$ is considered,
the whole process of querying for oracle $\phi$ costs only one time unit.
\begin{definition}
{\normalfont\cite{K91}}
Function $f:[a,b]\rightarrow \stdsetnameinform{R}$ is said to be
$\classnameinform{FPTIME}$ {\normalfont(}$\classnameinform{FEXPTIME}${\normalfont)}
computabel real function on $[a,b]$ if for all computable $x\in[a,b]$
function $\psi\in CF_{f(x)}$ ($\psi$ is from the definition of $CF$ computable
real function) is $\classnameinform{FPTIME}$
{\normalfont(}$\classnameinform{FEXPTIME}${\normalfont)} computable.
\end{definition}
Function $f$ computable at point $x\in[a,b]$ is defined in a similar way.

Set of polynomial time (exponential time) computable real functions on $[a,b]$ is denoted by
$$\classnameinform{FPTIME}_{C[a,b]}\ (\classnameinform{FEXPTIME}_{C[a,b]}).$$
%
%/////////////////////////////////////////////////////////////////////////////
%/////////////////////////////////////////////////////////////////////////////
%
\mysectionb
{Modulus function of uniform continuity}
{Modulus function of uniform continuity}
\begin{definition}
\label{Def:UCF}
{\normalfont\cite{K91}}
Let $f:[a,b]\to\stdsetnameinform{R}$ be a uniformly continuous function
on $[a,b]$. Function $\omega:\stdsetnameinform{N}\to\stdsetnameinform{N}$
is said to be a modulus function of uniform continuity of function $f$ on $[a,b]$ if
\begin{align*}
|x-y|\le 2^{-\omega(n)}\ \text{implies}\ |f(x)-f(y)|\le 2^{-n}
\end{align*}
for all $n\in\stdsetnameinform{N}$ and for all $x,y\in[a,b]$.
\end{definition}
In the following theorem it is said that $\omega$ is a computable function
for computable real functions (this theorem holds if a model of computable real functions is considered
such that real functions on $[a,b]$ are defined for all real numbers in $[a,b]$).
\begin{theorem}
{\normalfont\cite{K91}}
\label{Thr:CFFuncsModOFCont}
If function $f:[a,b]\to\stdsetnameinform{R}$ is a $CF$ computable real function on $[a,b]$ then there
exists a modulus function of uniform continuity of function $f$ on $[a,b]$.
\end{theorem}
It is said that there is a polynomial modulus function of uniform continuity of function $f$
on $[a,b]$ \cite{K91} if there exists a polynomial
$\omega:\stdsetnameinform{N}\to\stdsetnameinform{N}$ such that
\begin{align*}
|x-y|\le 2^{-\omega(n)}\ \text{implies}\ |f(x)-f(y)|\le 2^{-n}
\end{align*}
for all $n\in\stdsetnameinform{N}$ and for all $x,y\in[a,b]$.
\begin{theorem}
{\normalfont\cite{K91}}
If $f$ is a \classname{FPTIME} computable real function then there exists a polynomial modulus
function of uniform continuity of function $f$ on $[a,b]$; it means there exists a polynomial
$\omega:\stdsetnameinform{N}\to\stdsetnameinform{N}$ such that
\begin{align*}
|x-y|\le 2^{-\omega(n)}\ \text{implies}\ |f(x)-f(y)|\le 2^{-n}
\end{align*}
for all $n\in\stdsetnameinform{N}$ and for all $x,y\in[a,b]$.
\end{theorem}
The same results hold if we consider the evaluation of function $f$ at a point $x\in[a,b]$ and
consider the notion of modulus function of continuity at this point \cite{K84}.
%
%/////////////////////////////////////////////////////////////////////////////
%/////////////////////////////////////////////////////////////////////////////
%
\mysectiona
{Construction of computable real function $\mathcal{F}$}
{Construction of computable real function $\mathcal{F}$}
Let's construct computable real function $\mathcal{F}$ on $[0,1]$ such that
\begin{enumerate}
\item[1)]
{there does not exists polynomial modulus function of uniform continuity
of function $\mathcal{F}$ on $[0,1]$, and therefore function $\mathcal{F}$ is not a
polynomial time computable function on $[0,1]$ (moreover, there does not exist any
polynomial time modulus function of continuity of function $\mathcal{F}$ at any rational
point on $(0,1)$, so function $\mathcal{F}$ is not a polynomial time computable function
at any rational point on $(0,1)$),
}
\item[2)]
{there exists an exponential time algorithm for the evaluation of function $\mathcal{F}$
on $[0,1]$.
}
\end{enumerate}
The construction of function $\mathcal{F}$ on $[0,1]$ is as follows:
\begin{enumerate}
\item[1)]
{
let's define real function $\beta_{p,q}$ on $[0,1]$:
\begin{align*}
\beta_{p,q}(x)=
\begin{cases}
0 &
\text{if}\ 0 \le x \le \frac{p}{q},\\
2^q \left(x-\frac{p}{q}\right) &
\text{if}\ \frac{p}{q} \le x \le \frac{p+\frac{1}{2^q}}{q},\\
\frac{1}{q} &
\text{if}\ \frac{p+\frac{1}{2^q}}{q} \le x \le 1,
\end{cases}
\end{align*}
for natural numbers $p$ and $q$ such that $q\ge 1$ and $p\in [0..(q-1)]$;
}
\item[2)]
{                                           
let's define real function $\alpha_q$ on $[0,1]$:
\begin{align*}
\alpha_q(x)=\sum^{q-1}_{p=0}{\beta_{p,q}(x)};
\end{align*}
}
\item[3)]
{                                           
\begin{align}
\label{Eq:FSeries}
\mathcal{F}(x)=\sum^{\infty}_{q=1}{\frac{1}{q^2}\alpha_{q}(x)}.
\end{align}
}
\end{enumerate}
%
%/////////////////////////////////////////////////////////////////////////////
%/////////////////////////////////////////////////////////////////////////////
%
\mysectionb
{Function $\mathcal{F}$ is not polynomial time computable}
{Function $\mathcal{F}$ is not polynomial time computable}
Let's prove that function $\mathcal{F}$ has the following properties:
\begin{enumerate}
\item[1)]
{function $\mathcal{F}$ is a monotonically increasing function on $[0,1]$;}
\item[2)]
{function $\mathcal{F}$ is a uniformly continuos function on $[0,1]$;}
\item[3)]
{$\omega(n)\ge 2^{C\cdot n}$, wherein $\omega$ is the function from definition \ref{Def:UCF},
should hold if we evaluate function $\mathcal{F}$ in precision $2^{-n}$.}
\end{enumerate}
Functions $\beta_{p,q}$ are monotonically increasing functions, therfore 
$\alpha_q$ are monotonically increasing functions as sums of monotonically increasing functions
$\beta_{p,q}$, and function $\mathcal{F}$ is monotonically increasing function as a sum of
monotonically increasing functions $\alpha_q$. So, point 1) holds.

Because
\begin{align*}
|\alpha_q(x+\delta)-\alpha_q(x)|\le 1,
\end{align*}
the following holds for all real $x\in[0,1]$ and real $\delta$, wherein $\delta\le 1$:
\begin{align*}
|\mathcal{F}(x+\delta)-\mathcal{F}(x)|&\le
\sum_{q:\delta\le \frac{1}{q}}^{}{\frac{1}{q^2}}+
\sum_{q:\delta > \frac{1}{q}}^{}{\frac{1}{q^2}}\le
\frac{1}{\delta}+\delta.
\end{align*}
So, point 2) holds (taking into account point 1)).

Further, we have
\begin{enumerate}
\item[a)]
{functions $\beta_{p,q}$ and function $\mathcal{F}$ are monotonically increasing functions, and}
\item[b)]
{if $\delta\le\frac{1}{2^n}$, $n\le q\le n^2$, and $x=\frac{p}{q}$ then
\begin{align*}
x=\frac{p}{q}<\frac{p+\frac{1}{2^q}}{q}<x+\delta\qquad\text{and}\qquad
\left|\frac{1}{q}\right|\ge \left|\frac{1}{\log(\delta)^2}\right|.
\end{align*}
}
\end{enumerate}
Therefore, for all real $\delta$, wherein $\delta\le 1$, there exist $x$, namely
$x=\frac{p}{q}$, such that the following holds:
\begin{align*}
|\mathcal{F}(x+\delta)-\mathcal{F}(x)|&>
\frac{1}{q^2}\left|\beta_{p,q}(x+\delta)-\beta_{p,q}(x)\right|=
\frac{1}{q^2}\cdot\frac{1}{q}=
\left(\log_2\left(\delta\right)\right)^{-6}.
\end{align*}
Hence,
$$\left(\log(\delta)\right)^{-6}\le 2^{-n}$$ should hold
if we require that $$|\mathcal{F}(x+\delta)-\mathcal{F}(x)|\le 2^{-n}$$ holds;
it means $\omega(n)\ge 2^{C\cdot n}$ should hold for $\delta=2^{-\omega(n)}$.
So, point 3) holds.
                                                                                    
It follows from point 3) that proposition \ref{Prop:UnifContNotExist} holds.
\begin{proposition}
\label{Prop:UnifContNotExist}
There does not exist polynomial modulus function of uniform continuity of function $\mathcal{F}$
on $[0,1]$.
\end{proposition}
Let $x$ be rational $\frac{p}{q'}\in (0,1)$, real $\delta$ be $\frac{1}{2^n}$; in that case
point b) holds for $q=(q')^k$, such that $n\le (q')^k\le n^2$ for some natural number $k$, and
for $x=\frac{p\cdot (q')^{k-1}}{(q')^k}$; therefore, $\omega(n)\ge 2^{C\cdot n}$ should hold at $x$.
So, proposition \ref{Prop:ContNotExist} holds.
\begin{proposition}
\label{Prop:ContNotExist}
For each rational number $x\in(0,1)$ there does not exist polynomial modulus function of continuity of
function $\mathcal{F}$ at point $x$.
\end{proposition}
So, the following theorems hold.
\begin{theorem}
Real function $\mathcal{F}$ is not a polynomial time computable real function on $[0,1]$
of real numbers.
\end{theorem}
\begin{theorem}
For each rational number $x\in(0,1)$ real function $\mathcal{F}$ is not a polynomial time computable
real function at point $x$.
\end{theorem}
Let's note that function $\mathcal{F}$ is a nowhere-differentiable function on $[0,1]$.
%
%
%/////////////////////////////////////////////////////////////////////////////
%/////////////////////////////////////////////////////////////////////////////
%
\mysectionb
{Evaluation of function $\mathcal{F}$ in exponential time}
{Evaluation of function $\mathcal{F}$ in exponential time}
Because
\begin{align*}
|\alpha_q(x)|\le 1,
\end{align*}
for the remainder of series \eqref{Eq:FSeries} the following holds:
\begin{align*}
|R_t(x)|\le
\left|\sum^{\infty}_{q=t+1}{\frac{1}{q^2}}\right|<
C_1\cdot 2^{-C_2 \log(q)}.
\end{align*}
Therefore, if we sum $2^{C\cdot n}$ terms of the series and evaluate each of that terms
(which are functions $\alpha_q$) in precision $2^{-C\cdot n}$ then we evaluate series
\eqref{Eq:FSeries} in precision $2^{-n}$.

Because $\omega(n)$, wherein $\omega$ is the function from definition \ref{Def:UCF}, is
exponential in $n$, the time complexity of functions $\alpha_{q}$
is exponential in $n$; therefore, the time complexity of the evaluations of
$\mathcal{F}(x)$ in precision $2^{-n}$ is exponential in $n$.

To evaluate functions $\beta_{p,q}$ at points $\frac{p}{q}$ and $\frac{p+\frac{1}{2^q}}{q}$, let's use the
following result from \cite{K84}. Let
\begin{enumerate}
\item[1)]
{$f$ be a computable real function on $[a,b]$,}
\item[2)]
{$g$ be a computable real function on $[b,c]$, and}
\item[3)]
{$f(b)=g(b)$.}
\end{enumerate}
In that case, function $h$ defined by equation
\begin{align*}
h(x)=f(\max(x,z))+g(\max(x,z))-f(z)
\end{align*}
is a computable real function and has the following property:
\begin{align*}
h(x)=
\begin{cases}
f(x) & \text{if}\ x\in[a,b],\\
g(x) & \text{if}\ x\in[b,c].\\
\end{cases}
\end{align*}
\begin{theorem}
Real function $\mathcal{F}$ is an exponential time computable real function on $[0,1]$ of
real numbers.
\end{theorem}
%
%/////////////////////////////////////////////////////////////////////////////
%/////////////////////////////////////////////////////////////////////////////
%
\mysectiona
{Conclusion}
{Conclusion}
One of the open questions here is as follows: if function $\mathcal{F}$ is polynomial time computable
at irrational numbers $x\in(0,1)$ ? If not, how a function that is not polynomial time computable
at every real $x\in[0,1]$ could be constructed ?

It could as well be interesting to construct computable real functions that are computable in
a time complexity class, but not computable in a space complexity class, or that are computable in
a space complexity class, but not computable in a time complexity class, and so on.
%
%/////////////////////////////////////////////////////////////////////////////

%/////////////////////////////////////////////////////////////////////////////

\end{document}